\documentclass[nohyper,12pt,letterpaper]{JHEP}
\usepackage{epsfig}

\newfont{\frak}{eufm10 scaled 1200}

\newfont{\Bbb}{msbm10 scaled 1200}     
\newcommand{\mathbb}[1]{\mbox{\Bbb #1}}
\DeclareSymbolFont{AMSa}{U}{msa}{m}{n}
\DeclareSymbolFont{AMSb}{U}{msb}{m}{n}
\let\Box\relax
\DeclareMathSymbol{\Box}{\mathord}{AMSa}{"03}

\title{Quantum Determinism for Free Vector Bosons in 3 Dimensions}
\author{Philippe Pouliot\\
 Physics Department\\
University of Texas at Austin\\
Austin, TX 78712 USA\\
Email: \email{pouliot@physics.utexas.edu}}

\abstract{We apply 't Hooft's deterministic quantum mechanics approach to 
free vector bosons in three dimensions
and check Lorentz invariance.
This approach does not work for the conformal group, for free bosons in 
two dimensions. 
This presents a technical difficulty for constructing a ``deterministic 
string theory''.}

\keywords{String Theory, Deterministic Quantum Mechanics}

\preprint{UTTG-08-01}

\begin{document}

\section{Introduction}
In a series of papers, 't Hooft has advocated a deterministic approach to 
quantum mechanics. For a review, see \cite{'tHooft:2001fb}. We were 
motivated by his suggestion to use his procedure
for string theory (see bottom of page 6 of \cite{'tHooft:2001fb}). In 
\cite{'tHooft:2001ar}, 't Hooft introduces a new kind of (non-local) 
gauge symmetry. 
It removes the amplitude of the field momentum modes, keeping only their 
phases. As explained in \cite{'tHooft:2001ar},  the classical 
free boson field with this gauge symmetry is equivalent to a quantum free
 boson. Furthermore, except for the zero modes, the spectrum of each 
momentum mode can be 
truncated to a finite number of states.

In this note, we point out a technical difficulty in carrying out this 
program for
 string theory: the ``global gauge symmetry'' is not compatible with the 
conformal group. It is therefore not obvious how to follow through
with the general philosophy. We should quickly point out that there might 
be clever ways around this problem, and that 't Hooft's deterministic 
approach remains very interesting. 
 Our argument is very simple: 
a conformal transformation involves an arbitrary function $\epsilon(z)$ 
and it is not possible to reabsorb the effect of this transformation 
with a transformed kernel that is independent of the world-sheet space 
coordinate of the scalar field. 

In section 2, we review the Lorentz invariance of the scalar; in 
section 3, we show the Lorentz covariance of the free vector boson in 
three dimensions; in section 4, we
show that the approach works for dilatations for the free boson in two 
dimensions but not for other conformal transformations.

\section{Lorentz invariance}
The reader looking for new results should skip this section.
\subsection{Boosts}
As a warm-up, we review 't Hooft's check that a real scalar field remains 
invariant under Lorentz transformations if the kernels transform in a 
certain way \cite{'tHooft:2001ct}.
The purpose of presenting these details here is that the calculations of 
the later sections are very similar, so we will be able to skip some 
details later.
Consider a real scalar free field $\phi(x,t)$ of mass $\mu$ in $D+1$ 
dimensions, satisfying $\ddot\phi=(\triangle-\mu^2)\phi$. 
Under 't Hooft's gauge transformation, $\phi$ and $\dot\phi$ become:
\begin{eqnarray}
\phi^{(g)}(x,t) & = & \int d^Dy \  K_1(y) \phi(x+y,t) + K_2(y) 
\dot\phi(x+y,t) \nonumber \\
\dot \phi^{(g)}(x,t) & = & \int d^Dy\  K_1(y) \dot\phi(x+y,t) + 
K_2(y) (\triangle_x-\mu^2)\phi(x+y,t),
\end{eqnarray}
using the equations of motion to replace $\ddot\phi$ by 
$(\triangle-\mu^2)\phi$.
Are the gauge transformed $\phi^{(g)}$ and $\dot\phi^{(g)}$ 
still Lorentz invariant?
Under an infinitesimal boost in the $x_1$ direction, 
\begin{eqnarray}
\delta\phi(x,t) &  = &  x_1 \dot\phi(x,t) + t \partial_{x_1} 
\phi(x,t)  \nonumber \\ 
\label{Ltrans} \delta\dot\phi(x,t) & = & x_1 (\triangle-\mu^2) 
\phi(x,t) + t\partial_{x_1} \dot\phi(x,t) + \partial_{x_1} \phi(x,t).
\end{eqnarray}
We can set $t=0$ at this stage: if $\phi$ and $\dot\phi$ transform 
properly at $t=0$, they will remain so at later times by the 
equations of motion.
But we will keep $t$ explicitly to illustrate that such terms follow 
through straightforwardly.
If we assume that the kernels transform as follows under this boost:
\begin{eqnarray}
\delta K_1(y) & = & -y_1(\triangle_y-\mu^2) K_2(y) - 
\partial_{y_1} K_2(y) \nonumber \\
\delta K_2(y) & = & - y_1 K_1(y), \label{Lkernel}
\end{eqnarray}
then $\phi^{(g)}$ and $\dot\phi^{(g)}$ transform correctly:
\begin{eqnarray}
\delta\phi^{(g)}(x,t) &  = &  x_1 \dot\phi^{(g)}(x,t) + 
t \partial_{x_1} \phi^{(g)}(x,t) \nonumber  \\
\delta\dot\phi^{(g)}(x,t) & = & x_1 (\triangle-\mu^2) \phi^{(g)}(x,t) 
+ t\partial_{x_1} \dot\phi^{(g)}(x,t) + \partial_{x_1} \phi^{(g)}(x,t). 
\end{eqnarray}
To check that, just substitute equations \ref{Ltrans} and \ref{Lkernel} 
into:
\begin{eqnarray}
\delta\phi^{(g)}(x,t) & = & \int\ K_1 \delta\phi + (\delta K_1) \phi + 
 K_2 \delta\dot\phi + (\delta K_2) \dot\phi \nonumber \\
\delta\dot\phi^{(g)}(x,t) & = & \int \ K_1 \delta\dot\phi + 
(\delta K_1) \dot\phi +  K_2 \delta((\triangle-\mu^2)\phi) + 
(\delta K_2) (\triangle-\mu^2) \phi. 
\end{eqnarray}
Note that $\delta$ acts on $\triangle$ in the above equation:
\begin{equation}
\delta((\triangle-\mu^2)\phi)= (\triangle-\mu^2)\delta\phi+2 
\partial_{x_1}\dot\phi.
\end{equation}
For the reader's convenience, we give the details of the 
computation; substituting:
\begin{eqnarray}
\delta\phi^{(g)}(x,t) & = & \int d^Dy \  K_1(y) 
\left((x_1+y_1)\dot\phi(x+y)+t\partial_{x_1}\phi(x+y)\right) \nonumber \\
&-& \left(y_1 (\triangle_y-\mu^2)K_2(y)+\partial_{y_1}K_2(y)\right) 
\phi(x+y) \nonumber \\
& + &  K_2(y) \left( (x_1+y_1)(\triangle_x-\mu^2)\phi(x+y) + 
t \partial_{x_1}\dot\phi(x+y) + 
\partial_{x_1}\phi(x+y)\right) \nonumber \\
& -&  y_1 K_1(y) \dot\phi(x+y).  \nonumber 
\end{eqnarray}
The terms containing $K_1$ cancel in part to leave $x_1 
\int K_1\dot\phi +t\partial_{x_1} \int K_1 \phi$.
For the terms containing $K_2$, after integrations by parts 
and discarding the surface terms, so that all the derivatives 
are with respect to the variable $x$, we
get $ x_1 (\triangle_x-\mu^2) \int K_2 \phi + t \partial_{x_1} 
\int K_2\dot\phi$.
They combine to give:
\begin{eqnarray}
\delta\phi^{(g)}(x,t)& = & x_1 \int (K_1\dot\phi + 
K_2 (\triangle_x-\mu^2) \phi)  + t\partial_{x_1} 
\int (K_1\phi + K_2\dot\phi)\nonumber \\
& = & x_1 \dot\phi^{(g)} + t \partial_{x_1} \phi^{(g)}.
\end{eqnarray}
Similarly, 
\begin{eqnarray}
\delta\dot\phi^{(g)}(x,t) & = & \int d^Dy\ K_1(y) 
\left( (x_1+y_1)(\triangle_x-\mu^2)\phi(x+y) + 
t\partial_{x_1}\dot\phi(x+y) + \partial_{x_1}\phi(x+y) \right) \nonumber \\
&-& \left( y_1 (\triangle_y-\mu^2) K_2(y)  
+ \partial_{y_1} K_2(y) \right) \dot\phi(x+y) \nonumber \\
&+&  K_2(y)  (\triangle_{x+y}-\mu^2)
\left((x_1+y_1)\dot\phi(x+y)+ t\partial_{x_1}\phi(x+y)) \right) +
2 K_2(y)\partial_{x_1}\dot\phi  \nonumber \\
&-& y_1 K_1(y)  (\triangle_x-\mu^2) \phi(x+y). \nonumber \\
&=&  \left(x_1 (\triangle-\mu^2)+\partial_{x_1}\right) 
\int( K_1\phi+K_2\dot\phi) + t\partial_{x_1} 
\int(K_1\dot\phi+K_2(\triangle-\mu^2)\phi) \nonumber \\
&=&  x_1 (\triangle-\mu^2) \phi^{(g)}(x,t) + 
t\partial_{x_1} \dot\phi^{(g)}(x,t) + \partial_{x_1} \phi^{(g)}(x,t).
\end{eqnarray}

\subsection{Rotations and Translations}
Under translations, the kernels are invariant: $\delta K_{1,2}=0$.
Under rotations, the kernels transform as scalars. 
That is, for a rotation in the $1-2$ plane,
\begin{eqnarray}
\delta \phi(x,t) & = & (x_1 \partial_{x_2} - 
x_2 \partial_{x_1}) \phi(x,t) \nonumber \\
\delta K_{1,2}(y) & = & (y_1 \partial_{y_2}- 
y_2 \partial_{y_1}) K_{1,2}(y).
\end{eqnarray}
 
\section{The free vector boson in $2+1$ dimensions}
In \cite{'tHooft:2001ct}, it is shown how the quantum 
deterministic approach works for a free vector boson in 3+1 dimensions.
The free vector boson in $2+1$ dimensions is simpler, but not 
entirely trivial. The key is understanding what equation $5.3$ 
in \cite{'tHooft:2001ct} becomes.
Lorentz invariance works for the vector in $2+1$ dimensions 
because it can be dualized to a scalar. 
We do not know how to generalize these examples to more than 
four dimensions: because the little group is larger,
it is not clear how to write a decomposition such as 
equation 5.3 in \cite{'tHooft:2001ct}.

We now show how the vector in $2+1$ dimensions works in some detail.
We write the duality equation 
$F_{\mu\nu} = \epsilon_{\mu\nu\rho}\partial^\rho \phi$ in momentum space.
With the momentum in the direction $k_2=k$, we have  
$k_1=0$ and the frequency $\omega = |k|$.
In the gauge $A_0=0$, there is the constraint $\partial_i A^i=0$, which is 
$k_1 A_1 + k_2 A_2 = 0$, and so $A_2(k)=0$ too. 
\begin{eqnarray}
F_{01} &=& \partial_0 A_1 =  \partial_2\phi  \nonumber \\
F_{02} &=& \partial_0 A_2  =  -\partial_1\phi  \nonumber \\
F_{12}&=& -\partial_2 A_1 = -\partial_0 \phi. 
\end{eqnarray}
In momentum space:
\begin{eqnarray}
\dot A_1 &=& i k \phi \nonumber \\
\dot A_2 &=& 0 \nonumber \\
i k A_1 & = & \dot \phi. \label{Aphi}
\end{eqnarray}
Thus $A_2$ stays zero and $A_1$ satisfies its 
equation of motion: $\ddot A_1=i k\dot\phi=-k^2 A_1$.
We now want $\phi$, $\dot\phi$ to transform in the usual way.
In momentum space, this is \cite{'tHooft:2001ar}:
\begin{eqnarray}
\phi^{(g)} &=& K_1\phi+i K_2\dot\phi \nonumber\\
\dot\phi^{(g)} & =& K_1\dot\phi - i K_2 k^2 \phi. 
\end{eqnarray}
And therefore, by equation \ref{Aphi}, $A_1$ transforms as:
\begin{eqnarray}
A^{(g)}_1 &=& K_1 A_1 +i K_2 \dot A_1 \nonumber \\
\dot A^{(g)}_1 &=&  K_1 \dot A_1 - i k^2 K_2 A_1.   
\end{eqnarray}
The Lorentz invariance of $\phi^{(g)}$ will carry 
over to $A^{(g)}$. Let's see this explicitly.
Returning to coordinate space:
\begin{eqnarray}
A_1^{(g)}(x,t) &=& \int d^2y\ K_1(y) A_1(x+y,t) + 
K_2(y) \dot A_1(x+y,t) \nonumber \\
\dot A_1^{(g)}(x,t) &=& \int d^2y\ K_1(y) \dot 
A_1(x+y,t) + K_2(y)\triangle_x  A_1(x+y,t).
\end{eqnarray}
Since $\partial_i A_i=0$, we can rewrite it as: 
\begin{eqnarray}
A_i^{(g)}(x,t)&=& \int d^2y\ K_1(y) A_i(x+y,t)+L_1(y) 
\dot A_i(x+y,t) \nonumber \\
\dot A_i^{(g)}(x,t)&=& \int d^2y\ K_1(y)\dot A_i(x+y,t) + 
L_1(y) \triangle A_i(x+y,t),
\end{eqnarray}
where we relabeled $L_1=K_2$ to conform to the notation of  
\cite{'tHooft:2001ct}.

The rest of the proof of Lorentz invariance can be read off from  
\cite{'tHooft:2001ct} by setting
$K_2=L_2=0$ and dropping one coordinate.
\begin{eqnarray}
\delta A_1 &=& x_2 \dot A_1 + t\partial_2 A_1 +\partial_1 
\Lambda \nonumber \\
\delta A_2 &=& x_2 \dot A_2 + t\partial_2 A_2 +\partial_2 
\Lambda \nonumber \\
\delta A_0 &=& A_2 + \partial_0 \Lambda, 
\end{eqnarray}
so $\Lambda = -\triangle^{-1} \dot A_2$ and $\partial_0 \Lambda=-A_2$.
At $t=0$:
\begin{eqnarray}
\delta A_i&=& x_2\dot A_i -\triangle^{-1} \partial_i \dot A_2 \nonumber\\
\delta \dot A_i &=& x_2\triangle A_i + \partial_2 A_i -\partial_i A_2 
\end{eqnarray}
and
\begin{eqnarray}
\delta K_1(y) & = & -y_2\triangle_y L_1(y) - 
\partial_{y_2} L_1(y) \nonumber \\
\delta L_1(y) & = & - y_2 K_1(y). 
\end{eqnarray}

\section{Conformal Invariance}

We now consider whether 't Hooft's procedure is applicable 
to conformal invariance in two dimensions. We will find 
out that it is not directly applicable. 

We can work with world-sheet coordinates $z=\sigma+i\tau$ 
and $\bar z$, say on the cylinder. Under a conformal transformation, 
\begin{equation}
z\rightarrow z + \epsilon(z),  \label{conformal}
\end{equation}
where $\epsilon(z)$ is a holomorphic function.
The free world-sheet periodic boson $X(z,\bar z)$ satisfies 
$\partial\bar\partial X(z,\bar z)=0$, so it splits into 
left-movers $X_L(z)$ and right-movers $X_R(\bar z)$; 
$X(z,\bar z) = X_L(z)+X_R(\bar z)$; 
$\dot X = i(\partial X_L-\bar \partial X_R)$. 
It is sufficient to look at the left-movers for our argument.
Under \ref{conformal}, 
\begin{eqnarray}
\delta_{\epsilon,\bar\epsilon} \Phi & = & 
(h\partial\epsilon+\epsilon\partial + 
\bar h\bar\partial\bar\epsilon+\bar\epsilon\bar\partial)
\Phi(z,\bar z) \nonumber\\
\delta X_L(z) & = & \epsilon(z)\partial X_L(z) \nonumber \\
\delta \partial X_L(z)& =& \epsilon(z)\partial^2X_L(z)+
(\partial\epsilon(z))X_L(z). 
\end{eqnarray}
Under 't Hooft's gauge transformation, 
\begin{eqnarray}
X^{(g)}_L(z) & = &\int dy\ K_1(y) X(z+y) + 
i K_2(y)\partial_z X(z+y) \nonumber\\
i\partial_z X^{(g)}_L(z) & =&  
\int dy\ K_1(y) i\partial_z X(z+y) - K_2(y)\partial^2_z X(z+y), 
\end{eqnarray}
with $y$ a real, one-dimensional variable. 
The transformed $X^{(g)}_L$ and $\partial X^{(g)}_L$ are
still holomorphic. However, they do not transform 
properly under conformal transformations. 
For example, take $\epsilon(z)=\epsilon_n z^n$, 
for an integer $n>2$. Then
\begin{eqnarray}
\delta X^{(g)}_L &=& \int dy\ \delta K_1 X + 
i\delta K_2 \partial X \nonumber \\
 & + & K_1 \epsilon_n (  z+ y)^n \partial X + 
i K_2 \epsilon_n \left((z+y)^n\partial^2 X +
 n (z+y)^{n-1}\partial X \right) .
\end{eqnarray}
Expanding $(z+y)^n$ using the binomial theorem, 
we see that the $z^n$ term is what we want to keep 
for the Lorentz transformation of $X$.
However, the subleading terms $z^{n-k} y^k$ are 
$z$-dependent terms. Using integrations by parts, 
one can hope at best to remove two subleading powers of $z$ from 
the second-derivative $\partial^2_z$. The rest of the 
$z$ dependence remains, 
and therefore the variation of the kernels $K_{1,2}$ 
would have to depend on $z$, which is not allowed.

\subsection{Special conformal transformations}

For the global conformal group, the procedure also 
does not work for special conformal transformations ($n=2$).
Although the unwanted terms can be cancelled by 
variations of the kernels, these variations do not 
have the right symmetries under 
$y\rightarrow -y$.  We use a prime to denote a 
$\sigma$ derivative. There are two kinds of special 
conformal transformations.
The first type, at $\tau=0$:
\begin{eqnarray}
\delta X &=& \sigma^2 X' \nonumber \\
\delta \dot X &=& \sigma^2 \dot X' + 2 \sigma \dot X.
\end{eqnarray}
In $\delta X^{(g)}$, collect the terms without time derivatives:
\begin{equation}
\delta X^{(g)} = \sigma^2\partial_\sigma \int K_1 X + 
\left(\int \delta K_1 X - \partial_y (K_1 (\underline{\ y^2\ } + 
2 y \sigma)) X  \right) + \cdots
\end{equation}
The underlined term, $\partial_y( K_1 y^2) X$ is odd under 
$y\rightarrow -y$ and has the wrong symmetry property for a 
variation of $K_1$.
The second type:
\begin{eqnarray}
\delta X &=& \sigma^2 \dot X \nonumber \\
\delta \dot X &=& \sigma^2 X'' + 2\sigma X'.
\end{eqnarray}
In $\delta X^{(g)}$, collect the terms with time derivatives:
\begin{equation}
\delta X^{(g)} = \sigma^2 \int K_1 \dot X + 
\left(\int \delta K_2 \dot X + (K_1 (\underline{\ y^2\ } + 2 y \sigma)) 
\dot X  \right) + \cdots
\end{equation}
The underlined term, $K_1 y^2\dot X$ is even under 
$y\rightarrow -y$ and has the wrong symmetry property 
for a variation of $K_2$.

\subsection{Dilatations}
The procedure does finally work for dilatations (and of course boosts).
Under a dilatation at $\tau=0$,
\begin{eqnarray}
\delta X &=& \sigma X' \nonumber \\
\delta \dot X &=& \sigma \dot X'+\dot X.
\end{eqnarray}
The variations of the kernels are:
\begin{eqnarray}
\delta K_1(y) &=& (yK_1(y))' \nonumber \\
\delta K_2(y) &=&  yK_2'(y). 
\end{eqnarray}

\acknowledgments

This work was supported by NSF grant PHY-0071512.

\end{document}